\documentstyle[aps,preprint]{revtex}
\begin{document}
\draft
\tightenlines
 
\title{Path Integrals and Perturbation Theory for Stochastic Processes}
\author{Ronald Dickman$^\dagger$ and Ronaldo Vidigal}
\address{
Departamento de F\'{\i}sica, ICEx,
Universidade Federal de Minas Gerais,\\
30123-970
Belo Horizonte - MG, Brasil\\
}

\date{\today}

\maketitle
\begin{abstract}
We review and extend the formalism introduced by Peliti,
that maps a Markov process to a path-integral representation.
After developing the mapping, we apply it to some illustrative
examples: the simple decay process, the birth-and-death process,
and the Malthus-Verhulst process.  In the first two cases we show
how to obtain the exact probability generating function using the
path integral.  We show how to implement 
a diagrammatic perturbation theory for processes that do not 
admit an exact solution.  Analysis of a set of coupled Malthus-Verhulst
processes on a lattice leads, in the continuum limit, to a
field theory for directed percolation and allied models.
\end{abstract}
\vspace{1cm}

$^\dagger$ email: dickman@fisica.ufmg.br

\newpage
\section{Introduction}

It is often noted that nonequilibrium statistical mechanics lacks the
comprehensive formalism of ensembles that has proved so useful in
equilibrium.  The reason is that equilibrium statistical mechanics
treats stationary states for a special subclass of systems, that possess
detailed balance.  This allows one to bypass the dynamics, and study
stationary properties directly.  For systems out of 
equilibrium, we generally do not have such a shortcut, and must deal
with the full dynamical problem, even if our goal is only to obtain
stationary properties.  In the case of stochastic systems with a discrete
state space, the fundamental description is given by the master
equation, which governs the evolution of the probability distribution.
This class of problems includes a wide range of systems of current interest,
that exhibit phase transitions or scale invariance far from
equilibrium: driven lattice gases,
birth-and-death processes such as directed percolation
or the contact process, sandpile models, and interface growth models.

One of the more powerful tools for studying stochastic models
is a formalism that maps the process to a 
path-integral representation \cite{lss}.
The mapping generates an {\it effective action} that can be studied
using the tools of equilibrium statistical physics, for example,
the  renormalization group.  Several methods for mapping a
stochastic process to an equilibrium-like action have been proposed
\cite{msr,dedominicis,doi,grassb,janssen,cardy96,poisson}.  
In these notes we review the method developed by 
Peliti \cite{peliti85}, and apply it
to some simple stochastic processes.
This method has several advantages.  With it, one can map any birth-and-death
type process to a path-integral representation without ambiguity.
In particular, the step of writing a Langevin equation, and of
postulating noise autocorrelations, does not arise in this formalism.
Thus it provides a direct path from the model of interest
to an effective action, and (in the continuum limit), to its
field theory, without the uncertainties that often attend the 
specification of the noise term \cite{howard}.  
A second advantage, which we explore in some
examples, is that it leads to a systematic
perturbative analysis for Markov processes.

The principal aim of this article is to acquaint the reader 
with the formalism and provide a set of worked examples whose mastery
will allow one to apply the method to problems at the
frontier of research.  While Peliti's article \cite{peliti85} provides
an excellent exposition of the mapping, we include, for completeness,
a derivation of the central formulas.  Our development of the
perturbation theory differs somewhat from Peliti's. 
Most of the applications discussed are also new.

The balance of this article is structured as follows.  
In Sec. II we derive the path-integral representation, 
starting from the master equation.  Sec. III presents an application
to the simple decay process, and expressions for two-time joint
probabilities.  In Sec. IV we begin our discussion of diagrammatic
perturbation theory for the probability generating function, which is
illustrated with a pedagogical example.  This is extended in Sec. V
where we analyze the birth-and-death process using perturbation theory.
In Sec. VI a perturbation expansion for moments of the distribution
is developed, which turns out to be much simpler than that for the
full generating function.  This method is applied to the Malthus-Verhulst
process in Sec. VII.  In Sec. VIII we illustrate another application 
of the formalism, showing how the path-integral description for a lattice
of coupled Malthus-Verhulst processes leads, in the continuum limit,
to a field theory for directed percolation.  Sec. IX presents
a brief summary.

\section{From the master equation to a path integral}

In this section we recapitulate Peliti's derivation of the path
integral mapping.
We consider Markov processes in continuous time, and with a discrete
state space $n =0,1,2,...$.  (We may think of $n$ as the size of a certain
population.)  The probability $p_n(t)$ of state $n$ at time $t$
is governed by the master equation \cite{vanK,gardiner,tome}:

\begin{equation}
\frac{d p_n(t)}{dt} = -p_n(t) \sum_m w_{mn} + \sum_m w_{nm} p_m (t)  \;,
\label{meq}
\end{equation}
where $w_{mn} $ is the rate for transitions from $n$ to $m$.
(We study stationary {\it stationary} Markov processes, i.e.,
time-independent transition rates.)

We now associate a vector $|n\rangle$ in a Hilbert space with each
state $n$, \cite{doi} and for convenience define the inner product so:

\begin{equation}
\langle m|n\rangle = n! \delta_{m,n}  \;.
\label{ip}
\end{equation}
The identity may then be written as

\begin{equation}
1 = \sum_n \frac{1}{n!} |n \rangle \langle n|  \;.
\label{ident}
\end{equation}
(All sums run from zero to infinity unless otherwise specified.
Note that while we make use of many pieces of notation familiar
from quantum mechanics, we are dealing with c-number functions.) 
A probability distribution $\phi_n$ ($\phi_n \geq 0$, $\sum_n \phi_n = 1$),
may be represented as a linear combination of basis states:

\begin{equation}
|\phi \rangle = \sum_n \phi_n |n \rangle   \;.
\label{state}
\end{equation}

The Hilbert space formalism is useful because it provides a simple
way to express the evolution in terms of creation ($\pi$) and
annihilation ($a$) operators, which we define via:

\begin{equation}
a |n \rangle = n|n\!-\!1 \rangle  \;\;\;\;\;\;\;\;
\pi |n \rangle = |n\!+\!1 \rangle  \;.
\label{ops}
\end{equation}
These relations imply that $[a,\pi] = 1$.
Here again we note a fundamental difference from quantum mechanics:
expected values are {\it linear}, not bilinear, in $|\phi \rangle$.
The mean population size, for instance, is given by

\begin{equation}
{\sf E} [n(t)] = \langle \;| a |\phi(t) \rangle \;,
\label{expect}
\end{equation}
where 

\begin{equation}
\langle \;| \equiv \sum_m \frac{1}{m!} \langle m |
\label{nbra}
\end{equation}
is the the projection onto all possible states.

Central to our analysis will be the {\it probability generating
function} (PGF),

\begin{equation}
\Phi_t(z) \equiv \sum_n p_n(t) z^n  \;.
\label{pgf}
\end{equation}
We denote the PGF corresponding to state $|\phi \rangle $ as
$\phi(z) = \sum_n \phi_n z^n$.  (Note that $\phi(1) = 1$ by
normalization.)

Next consider the inner product between states
$|\phi \rangle $ and $|\psi \rangle $:

\begin{equation}
\langle \phi | \psi \rangle
= \sum_n \frac{1}{n!} \langle \phi | n\rangle \langle n | \psi \rangle
= \sum_n n! \phi_n \psi_n     \;.
\label{ip2}
\end{equation}
We can write this in terms of the corresponding PGFs if we note the identity

\begin{equation}
\int dz z^n \left( - \frac{d}{dz} \right)^m \delta(z) = n! \delta_{n,m} \;,
\label{ident1}
\end{equation}
which is readily proved, integrating by parts.
(Unless otherwise specified, all integrals are over the real axis.)
Then we have

\begin{eqnarray}
\nonumber
\langle \phi | \psi \rangle &=& 
\int dz \phi(z) \; \psi \left( - \frac{d}{dz} \right) \delta(z)
\\ 
&=& \int \frac{dz dz'}{2 \pi} \phi(z) \psi(iz') e^{-izz'}  \;,
\label{ip3}
\end{eqnarray}
where we used the integral representation of the $\delta$ function.

For birth-and-death processes, it is always possible to write the
master equation in terms of an evolution operator $L$ composed of creation
and annihilation operators (specific examples are considered below).
The master equation then takes the form

\begin{equation}
\frac{d |\phi \rangle}{dt} = L |\phi \rangle \;,
\label{meop}
\end{equation}
and has the formal solution

\begin{equation}
|\phi (t)\rangle = e^{Lt} |\phi (0) \rangle \equiv U_t |\phi (0) \rangle \;.
\label{fsol}
\end{equation}
This evolution has its analog in the space of probability generating
functions; it is for the analog of the operator $U_t$ in the PGF
representation that we shall develop a path-integral expression.  To do this
we define, for any operator $A$ in the Hilbert space, a function called
its kernel:

\begin{equation}
A(z,\zeta) =  \sum_{m,n} \frac{z^m \zeta^n}{m! n!} A_{m,n}
\label{kernel}
\end{equation}
where $A_{m,n} = \langle m |A| n\rangle$ are the matrix elements of $A$.
Suppose $|\psi \rangle = A |\phi \rangle $.  The PGF corresponding to
$|\psi \rangle $ is

\begin{eqnarray}
\nonumber
\psi (z)  &=& \sum_n \psi_n z^n = \sum_n \frac{z^n}{n!}
\langle n |A| \phi \rangle
\\ 
&=&
\nonumber
 \sum_{n,m} \frac{z^n}{n!} \langle n |A| m\rangle  \frac{1}{m!}
\langle m | \phi \rangle
\\ \nonumber
&=&
 \sum_{n,m} \frac{z^n}{n!} A_{n,m} \frac{1}{m!}
\int \frac{d\zeta d\zeta'}{2 \pi} \zeta^m \phi(i\zeta')
 e^{-i\zeta \zeta'}
\\ 
&=&
\int \frac{d\zeta d\zeta'}{2 \pi} A(z,\zeta) \phi(i\zeta')
 e^{-i\zeta \zeta'}    \;.
\label{pgfz}
\end{eqnarray}
We shall also require an expression for the kernel of a product of a pair of
operators, $A$ and $B$:

\begin{eqnarray}
\nonumber
AB (z,\zeta)  &=& \sum_{m,n} \frac{z^m}{m!} [AB]_{m,n}   \frac{\zeta^n}{n!}
\\ \nonumber
&=&
 \sum_{n,m,r,s} \frac{z^m}{m!} \langle m |A| r\rangle  \langle s |B| n\rangle
  \frac{\zeta^n}{n!} \frac{\delta_{r,s}}{r!}
\\ \nonumber
&=&
 \sum_{n,m,r,s} \int d\eta \frac{z^m}{m!} A_{m,r}B_{s,n}
  \frac{\zeta^n}{n!} \frac{1}{(r!)^2}
  \eta^r \left( - \frac{d}{d\eta} \right)^s \delta(\eta)
\\ \nonumber
&=&
\int \frac{d\eta d\eta'}{2 \pi} \sum_{n,s} A(z,\eta)  B_{s,n}
\frac{\zeta^n}{n!} \frac{(i\eta')^s}{s!} (i\eta')^s
 e^{-i\eta \eta'}    
\\ 
&=&
\int \frac{d\eta d\eta'}{2 \pi} A(z,\eta)  B(i\eta',\zeta)
e^{-i\eta \eta'}    \;.
\label{kerpr}
\end{eqnarray}

Given an operator $A$, we may put it in {\it normal form} by 
commuting all creation operators to the left of all annihilation operators;
it will then have the form:

\begin{equation}
A = \sum_{m,n} {\cal A}_{m,n} \pi^m a^n  \;.
\label{normal}
\end{equation}
With this we associate the {\it normal kernel}:

\begin{equation}
{\cal A}(z,\zeta) = \sum_{m,n} {\cal A}_{m,n} z^m \zeta^n  \;.
\label{knormal}
\end{equation}
The ordinary and normal kernels are related via:

\begin{equation}
A(z,\zeta) = e^{z\zeta} {\cal A}(z,\zeta) \;.
\label{kerrel}
\end{equation}
[To prove Eq.(\ref{kerrel}) we show that the coefficients of $z^m \zeta^m$
on the right- and left-hand sides are equal.  The coefficient on the
r.h.s. is:

\[
\sum_{r=0}^{\min[m,n]} \frac{1}{r!} {\cal A}_{m-r,n-r}   \;,
\]
while on the l.h.s. it is simply $A_{m,n}/(m!n!)$.  The matrix element,
however, may be written so:

\begin{eqnarray*}
A_{m,n} &=&  \sum_{r,s} {\cal A}_{r,s} \langle m| \pi^r a^s |n \rangle
\\
 &=&  \sum_r \sum_{s=0}^n {\cal A}_{r,s} \frac{n!}{(n\!-\!s)!}
 \langle m|n\!+\!r\!-\!s \rangle
\\
 &=&  \sum_{s=0}^n {\cal A}_{m-n+s,s} \frac{m!n!}{(n\!-\!s)!}
=  \sum_{t=0}^{\min[m,n]} {\cal A}_{m-t,n-t} \frac{m!n!}{t!}
\end{eqnarray*}
which establishes the identity.]

We may now develop a path-integral representation for $U_t(z,\zeta)$.
To begin, recall Trotter's formula, which allows us to write:

\begin{equation}
U_t = e^{tL} = \lim_{N \to \infty} \left( 1 + \frac{tL}{N} \right)^N \;.
\label{trotter}
\end{equation}
Each factor in the product has a corresponding kernel, which, using
Eq. (\ref{kerrel}), can be written

\begin{equation}
\left( 1 + \frac{tL}{N} \right) (z,\zeta)
= e^{z \zeta} \left( 1 + \frac{t}{N} {\cal L}  (z,\zeta) \right)  \;,
\label{kerU}
\end{equation}
with ${\cal L}  (z,\zeta)$ the normal kernel of the evolution operator.
Now using Eq. (\ref{kerpr}), we have

\begin{equation}
U_t (z,\zeta) = \lim_{N \to \infty} \prod_{j=1}^{N-1} \left( \int 
\frac{d\eta_j d\eta_j'}{2 \pi} e^{-i\eta_j \eta_j'} \right)
\prod_{k=1}^N \left\{
e^{i\eta_k' \eta_{k-1}} \left[ 1 + 
\frac{t}{N} {\cal L} (i\eta_k',\eta_{k-1})\right] \right\}     \;,
\label{Uexp1}
\end{equation}
or, rearranging,

\begin{eqnarray}
\nonumber
U_t (z,\zeta) = \lim_{N \to \infty} \prod_{j=1}^{N-1}  \int 
\frac{d\eta_j d\eta_j'}{2 \pi} &\!& \!\!\! \exp \left\{
\sum_{k=1}^{N-1} \left[ -i\eta_k' (\eta_k \!-\! \eta_{k-1})
+ \frac{t}{N} {\cal L} (i\eta_k',\eta_{k-1})\right]  \right.
\\ 
&+& \left.  
\frac{t}{N} {\cal L} (z,\eta_{N-1}) + z \eta_{N-1} \right\}   \;.
\label{Uexp2}
\end{eqnarray}
Finally, we let $N \to \infty$ with $t' = (k/N)t$, 
$\eta_k \to \psi(t') $ and $\eta_k' \to \psi'(t') $, to obtain 
a path-integral expression for $U_t (z,\zeta)$:

\begin{equation}
U_t (z,\zeta) = \int {\cal D} \psi {\cal D} \psi'
\exp \left\{ \!-\!\int_0^t dt'
 \left[ i\psi'(t') \dot{\psi}(t')
- {\cal L} (i\psi',\psi) \right]  
+ z \psi(t) \right\}   \;,
\label{ut}
\end{equation}
where the dot denotes a time derivative.
The functional integrals over
$\psi(s)$ and $\psi'(s)$ are for $0 < s < t$,
with boundary conditions $\psi(0) = \zeta$ and $i \psi'(t) = z$;
$\psi$ and $\psi'$ are real.   
The symbol $\int {\cal D} \psi$ is defined by the 
limiting process in Eq. (\ref{Uexp2}).
[The reader may wonder at this point what has become of the
factors of $2 \pi$ in the denominator of Eq. (\ref{Uexp2}).
The answer is that the prefactor, which for the moment is
undefined, will be fixed via normalization.]
Note also that the first term in the argument of the exponential  
could be written more precisely as $i\psi'(t') \dot{\psi}(t'^{-})$, 
i.e., the time derivative is evaluated at $t'-\epsilon$ with
$\epsilon \to 0$ from above.
While the function $\psi'(t')$ has no obvious physical 
significance, we shall see that $\psi$ is closely related to the
random variable $n(t)$ in the birth-and-death process.

The kernel $U_t (z,\zeta)$ has two principal uses.  First, from
Eqs. (\ref{pgfz}) and (\ref{kerpr}), we see that
$U_t$ provides a mapping between PGFs
at different times:

\begin{equation}
\Phi_t(z) = \int \frac{d\zeta d\zeta'}{2\pi} e^{-i\zeta\zeta'}
U_t(z,\zeta) \Phi_0(i\zeta')  \;\;(t\geq 0) .
\label{map}
\end{equation}
We make considerable use of this relation in the examples
that follow.  Evaluating the integral is particularly simple
if the initial distribution is Poisson.  Then $\Phi_0(z) = e^{p(z-1)}$,
and

\begin{equation}
\Phi_t(z) = e^{-p} U_t(z,p)  \; .
\label{mappoi}
\end{equation}
Another simple case is $p_n(0) = \delta_{n,n_0}$ (exactly $n_0$
individuals initially), corresponding to $\Phi_0(z) = z^{n_0}$,
which yields

\begin{equation}
\Phi_t(z) = \left. \frac{\partial^{n_0} U_t(z,\zeta)}  
{\partial \zeta^{n_0}} \right|_{\zeta=0}
\; .
\label{mapn0}
\end{equation}
Setting $z \!=\! 0$ in the above expressions, we obtain the probability
of the state $n \!=\! 0$, which in many cases is absorbing, so
that the survival probability is given by $P_s (t) = 1 \!-\! p_0 (t)$.

Independent of its probability interpretation, the
kernel has a second utility: we may treat the argument of the
exponential as an effective action.  In the continuum limit
of a system with many degrees of freedom,
the function $\psi(t)$ becomes the classical field
$\psi({\bf x},t)$ (similarly for $\psi'(t)$), leading to a
field theory for a Markov process originally described by transition
rates for particles on a lattice.  
(An example of this, for the problem of directed percolation,
is discussed in Sec. VIII.)
With such a field theory in hand,
we can apply methods such as the renormalization group to study
critical behavior.  The effective action is known once we construct
the evolution operator $L$; at no point do we need to write a Langevin
equation or stipulate noise properties.

\section{Decay Process}

As a simple example we consider exponential decay,
i.e., the Markov process with transition rates
$w_{m,n} = wn \delta_{m,n-1}$.  The evolution operator is

\begin{equation}
L = w(1-\pi)a \;.
\label{opexpdec}
\end{equation}
Since this is in normal order, we have

\begin{equation}
{\cal L} = w(1-i\psi')\psi
\end{equation}
and thus, 

\begin{equation}
U_t(z,\zeta)\! = \!\! \int  \!{\cal D} \! {\psi}
\! \int \! {\cal D} \psi'
 \exp \left[\! -\int_0^t\! \!dt'
 \left( i\psi' (t') \dot{ \psi} (t')
+  w(i\psi'-1)\psi 
 \right) + z\psi(t)\right] \;.
\label{ut1}
\end{equation}
It is convenient to transform away the linear term in the action
via the change of variable $i \hat{\psi} = i\psi' - 1$.  The 
first term in the integral then contributes the additional (boundary)
terms $-\psi(t) + \psi(0)$, and using $\psi(0) = \zeta$ we have

\begin{equation}
U_t(z,\zeta)\! = \!\! \int  \!{\cal D} \!
{\psi}
\! \int \! {\cal D} \hat{\psi}
 \exp \left[\! -\int_0^t\! \!dt'
i\hat{\psi}[\partial_t +w] \psi
+ (z-1)\psi(t) + \zeta \right] .
\label{ut1a}
\end{equation}
This can be evaluated exactly.  Recalling that
$\int d\omega e^{i\omega z} = 2\pi \delta(z)$ ,
we see that the functional integral

\[
\int \! {\cal D} \hat{\psi} \exp \left[\! -\int_0^t\! \!dt'
i\hat{\psi}[\partial_t +w] \psi \right]
\]
imposes the condition 

\begin{equation}
\frac{d \psi(s)}{ds} = - w\psi(s)
\label{ode}
\end{equation}
for $0 < s \leq t$, and so $\psi(t) = \psi(0) e^{-wt} = \zeta e^{-wt}$.
In other words the functional integral over $ \hat{\psi}$ yields a
product of $\delta$-functions:

\begin{equation}
\int \! {\cal D} \hat{\psi} \exp \left[\! -\int_0^t\! \!dt'
i\hat{\psi}[\partial_t +w] \psi \right]
= {\mbox const.} \times \prod_{0<s \leq t} \delta(\psi(s)-\zeta e^{-ws}),
\label{delta}
\end{equation}
which when inserted in Eq. (\ref{ut1a}) gives

\begin{equation}
U_t(z,\zeta)\! = {\cal C}
\exp \{\zeta[1+(z-1)e^{-wt}]\},
\label{ut1b}
\end{equation}
where ${\cal C}$ represents the as yet undetermined 
normalization factor.
Using this in Eq. (\ref{map}) results in

\begin{eqnarray}
\nonumber 
\Phi_t(z) &=& {\cal C}\int \frac{d\zeta d\zeta'}{2\pi} 
\exp \{i\zeta[-\zeta' -i(1 + (z-1)e^{-wt})]\} 
\Phi_0(i\zeta') 
\\
&=& {\cal C} \int d \zeta' \delta[i\zeta' - (1+(z-1)e^{-wt})]
\Phi_0(i\zeta')  
\\
&=& \Phi_0 [1+(z-1)e^{-wt}] ,
\end{eqnarray}
where in the final line we set ${\cal C} = 1$ to satisfy the
normalization condition, $\Phi(1) = 1$.
If there are exactly $n$ particles at time zero, $\Phi_0(z) = z^n$,
and 

\begin{equation}
\Phi_t(z) = [ze^{-wt} + 1-e^{-wt}]^n
\end{equation}
which on expanding yields

\begin{equation}
\phi_m(t) =
\left( \begin{array} {c}
 n \\
 m
        \end{array} \right) 
e^{-mwt}(1-e^{-wt})^{n-m}
\label{binom}
\end{equation}
as expected.

For a Poisson initial distribution we find

\begin{equation}
\Phi_t(z) = e^{p(z-1) e^{-wt}}
\end{equation}
corresponding to a Poisson distribution whose mean decays
exponentially: $p(t) = pe^{-wt}$.

\subsection{Joint Probabilities}

It is useful to extend the formalism to joint probabilities, i.e.,
for the values of the process at different times.  
For $t_1 \geq t_2$,
let $P(n_1,t_1;n_2,t_2|n_0,0)$ be the probability of state $n_1$ at time $t_1$
{\it and} $n_2$ at $t_2$, {\it given} $n_0$ at time 0.  The generating function 
for the joint probability is

\begin{eqnarray}
\Phi(z_1,t_1;z_2,t_2|n_0) &=& \sum_{n_1,n_2} z_1^{n_1} z_2^{n_2}
P(n_1,t_1;n_2,t_2|n_0,0)
 \nonumber \\
&=& \sum_{n_1,n_2} z_1^{n_1} z_2^{n_2}
P(n_1,t_1|n_2,t_2)P(n_2,t_2|n_0,0)
 \nonumber \\
&=& 
\sum_{n_1,n_2} z_1^{n_1} z_2^{n_2}
P(n_1,t_1-t_2|n_2,0)P(n_2,t_2|n_0,0) \;,
\end{eqnarray}
where in the second line we used the Markov property and in the third,
stationarity.  

Let us modify slightly our notation for the one-time
generating function, to include the initial condition,
$P(n,0) = \phi_n (0) = \delta_{n,n_0}$:

\begin{equation}
\Phi (z,t|n_0) = \sum_n z^n P(n,t|n_0,0) \;.
\end{equation}
Then we have

\begin{eqnarray}
\Phi(z_1,t_1;z_2,t_2|n_0) &=& \sum_{n_2} \Phi (z_1,t_1-t_2|n_2) \; z_2^{n_2}
P(n_2,t_2|n_0,0)
\nonumber \\
&=& \sum_{n_2} z_2^{n_2} P(n_2,t_2|n_0,0)
\int \frac{d\zeta d\zeta'}{2\pi} e^{-i\zeta\zeta'}
U_{t_1-t_2}(z_1,\zeta) \Phi(i\zeta',0|n_2) \;
\end{eqnarray}
where, in the second line, we used Eq. (\ref{map}).
Now, noting that $\Phi(i\zeta',0|n_2) = (i\zeta')^{n_2}$, we have

\begin{eqnarray}
\Phi(z_1,t_1;z_2,t_2|n_0) &=& \sum_{n_2} P(n_2,t_2|n_0,0)
\int \frac{d\zeta d\zeta'}{2\pi} e^{-i\zeta\zeta'}
(iz_2\zeta')^{n_2} U_{t_1-t_2}(z_1,\zeta) 
\nonumber \\
&=& 
\int \frac{d\zeta d\zeta'}{2\pi} e^{-i\zeta\zeta'}
U_{t_1-t_2}(z_1,\zeta) \Phi(iz_2\zeta',t_2|n_0)
\nonumber \\
&=& 
\int \frac{d\zeta d\zeta'}{2\pi} \int \frac{d\xi d\xi'}{2\pi} 
e^{-i\zeta\zeta'} U_{t_1-t_2}(z_1,\zeta) 
e^{-i\xi\xi'} U_{t_2}(iz_2\zeta',\xi) (i\xi')^{n_0} \;.
\label{2time}
\end{eqnarray}
Thus we have a formula analogous to Eq. (\ref{map}), for the two-time
generating function.  The generalization to $n$ times is straightforward.

For the decay process, Eq. (\ref{2time}) reads:

\begin{eqnarray}
\Phi(z_1,t_1;z_2,t_2|n_0) &=&
\int \frac{d\zeta d\zeta'}{2\pi} \int \frac{d\xi d\xi'}{2\pi} 
e^{-i\zeta\zeta'} \exp\left[\zeta\{(z_1-1)e^{-w(t_1-t_2)} + 1\}\right]
\nonumber \\
&\times& 
e^{-i\xi\xi'} \exp\left[\xi\{(iz_2\zeta'-1)e^{-wt_2} + 1\}\right] 
(i\xi')^{n_0} \;,
\end{eqnarray}
which, after integrations over $\delta$-functions, yields:

\begin{equation}
\Phi(z_1,t_1;z_2,t_2|n_0) = \left\{1 + e^{-wt_2}
\left[z_2 \left((z_1\!-\!1)e^{-w(t_1-t_2)} + 1\right) -1 \right] \right\}^{n_0} \;.
\end{equation}
From this we readily obtain the expectation:

\begin{equation}
\langle n_i \rangle = \frac{\partial \Phi (z_1,t_1;z_2,t_2|n_0)}
{\partial z_i}\left|_{z_1=z_2=1} \right. = n_0 e^{-wt_i}  ,
\end{equation}
and
\begin{equation}
\langle n_1 n_2 \rangle = \frac{\partial^2 \Phi (z_1,t_1;z_2,t_2|n_0)}
{\partial z_1 \partial z_2}|_{z_1=z_2=1} 
= n_0 e^{-wt_1} +  n_0(n_0\!-\!1)e^{-w(t_1+t_2)} \;.
\end{equation}
The covariance is then
\begin{equation}
\langle\langle n_1 n_2 \rangle\rangle = n_0 e^{-wt_1} [1-e^{-wt_2}] \;.
\end{equation}
For $t_1 = t_2$ we find $\langle\langle n(t) \rangle\rangle = n_0 e^{-wt} [1-e^{-wt}] $,
which is the variance of a binomial random variable [see Eq. (\ref{binom})].

\section{Perturbation Theory}

The preceding example illustrates the general procedure for calculating
probabilities, but no perturbative treatment was needed as the
action was purely bilinear in the fields.  An example of perturbation
theory is afforded by analyzing the Malthus-Verhulst process, in
which each individual has a rate $\lambda$ to reproduce, and a rate
of $\mu + \nu (n-1)$ to die, if the total population is $n$.
(The term $\propto \nu$ represents saturation, so that the population
does not grow without limit even if $\lambda > \mu$.)  For this
process the kernel is

\begin{equation}
{\cal L}(z,\zeta) = \lambda(z-1)z\zeta + \mu (1-z)\zeta 
+ \nu (1-z)z\zeta^2 ,
\label{lmv}
\end{equation}
since the corresponding evolution operator has the property

\begin{equation}
L|n\rangle = \lambda n [|n+1\rangle -|n\rangle] 
+ \mu n[|n-1\rangle -|n\rangle] 
+\nu n(n-1)[|n-1\rangle -|n\rangle],
\label{opmv}
\end{equation}
which evidently reproduces the rates defining the process.

Using Eq. (\ref{map}) we have (with $\overline{\psi} \equiv i\psi'$),

\begin{eqnarray}
\nonumber
U_t(z,\zeta)\! = \!\! \int  \!{\cal D} \! {\psi}
\! \int \! {\cal D} \overline{\psi}
\! \exp &\!& \!\!\!\!\! \left[\! -\!\int_0^t\! \!dt'
 \left( \overline{\psi} \dot{ \psi}
- \lambda (\overline{\psi}\!-\!1)\overline{\psi} \psi\! \!
- \! \mu(1\!-\!\overline{\psi})\psi \right. \right.
\\
&-& 
\left. \left. \nu (1\!-\!\overline{\psi})\overline{\psi} \psi^2
 \right)\! \!+ z\psi(t)\right] .
\label{ut2}
\end{eqnarray}
As before, we eliminate the linear term by changing variables,
$\hat{\psi} \!=\! i\psi' \!-\!1$, yielding

\begin{eqnarray}
\nonumber
U_t(z,\zeta)\! = \!\! \int  \!{\cal D} \! {\psi}
\! \int \! {\cal D} \hat{\psi}
 \exp \!\left[\! -\!\! \int_0^t\! \!dt'
 \left( \hat{\psi}[\partial_{t'} \!+\! w]\psi
 \right. \right.
\! &-& \! \left. \lambda \hat{\psi}^2 \psi   
\!+\! \nu \hat{\psi}(1\!+\!\hat{\psi}) \psi^2
 \right)
\\ 
 &+&\! \left. \zeta + (z\!-\!1)\psi(t)\right] ,
\label{ut2a}
\end{eqnarray}
where $w = \mu - \lambda$.  Separating the bilinear part of the
action from the cubic and quartic terms, we have

\begin{equation}
U_t(z,\zeta)\! = \!\! \int  \!{\cal D} \! {\psi}
\! \int \! {\cal D} \hat{\psi}
 \exp \!\left[\! -\! \int_0^t\! \!dt'
\hat{\psi}[\partial_{t'} \!+\! w]\psi
+\! \zeta + (z\!-\!1)\psi(t)\right]
\exp[-S_I],
\label{ut2b}
\end{equation}
where
\begin{equation}
S_I = \int_0^t dt'[- \lambda \hat{\psi}^2 \psi   
\!+\! \nu \hat{\psi}(1\!+\!\hat{\psi}) \psi^2 ] 
\equiv \int_0^t dt'{\cal L}_I (t').
\label{si}
\end{equation}

Evidently $S_I$ must be treated perturbatively.  It is interesting to note
that while the process with $\nu = 0$ admits an exact solution, the
cubic term generates a perturbation series in the present formalism.
When we expand $e^{-S_I}$ we obtain a series of functional integrals over
$\psi$ and $\hat{\psi}$ of various products of these fields times the
``Gaussian" factor $e^{-S_0}$.  
Consider the basic contraction:

\begin{equation}
[\psi(t_1) \hat{\psi}(t_2)] = \int  \!{\cal D} \! {\psi}
\! \int \! {\cal D} \hat{\psi}  \psi(t_1) \hat{\psi}(t_2)
 \exp \!\left[\! -\! \int_0^t\! \!dt'
\hat{\psi}[\partial_{t'} \!+\! w]\psi
+\! \zeta + (z\!-\!1)\psi(t)\right]  .
\label{brac}
\end{equation}
Let

\begin{eqnarray}
{\cal F}[\psi,\hat{\psi}] &=& 
 \exp \!\left[\! -\! \int_0^t\! \!dt'
\hat{\psi}[\partial_{t'} \!+\! w]\psi
+\! \zeta + (z\!-\!1)\psi(t)\right] \nonumber  \\
&=& 
\exp \!\left[\! -\!\int_0^t\! \!dt'
\psi[-\partial_{t'} \!+\! w]\hat{\psi} 
+ \zeta[\hat{\psi}(0) +1]\right]  ,
\end{eqnarray}
where in the second line we integrated by parts and used
the boundary conditions $\psi(0) = \zeta$ and $\hat{\psi}(t) = z\!-\!1$.
Since the contraction $[\psi(t_1) \hat{\psi}(t_2)]$ is the functional
integral of $\psi(t_1) \hat{\psi}(t_2) {\cal F}$, it is convenient to
introduce the operator ${\cal K}$ via the relation

\begin{equation}
{\cal K}(t_2) {\cal F}[\psi,\hat{\psi}]
= \hat{\psi}(t_2) {\cal F}[\psi,\hat{\psi}].
\label{defk}
\end{equation}
The fact that 

\begin{equation}
\frac{\delta}{\delta \psi(t'')} 
{\cal F}[\psi,\hat{\psi}]
= -(-\partial_{t''} + w) \hat{\psi}(t'')  {\cal F}[\psi,\hat{\psi}],
\label{fderiv}
\end{equation}
suggests that ${\cal K}$ take the form

\begin{equation}
{\cal K}(t_2) = \int_0^t dt'' \kappa(t'',t_2) \frac{\delta}{\delta \psi(t'')} + {\cal B}
\label{k}
\end{equation}
where ${\cal B}$ is a boundary term.  Using Eq. (\ref{fderiv}) and integrating
by parts yields

\begin{eqnarray}
\int_0^t dt''&\!& \!\!\!\!\!\! \kappa(t'',t_2) \frac{\delta}{\delta \psi(t'')}
{\cal F}  \nonumber  \\
&=& \left[ -\int_0^t dt'' \hat{\psi}(t'') (\partial_{t''} + w) \kappa(t'',t_2)
+(z\!-\!1)\kappa(t,t_2) - \hat{\psi}(0) \kappa(0,t_2) \right] {\cal F}.
\end{eqnarray}
Consistency with Eq. (\ref{defk}) then requires that

\begin{equation}
(\partial_{t''} + w) \kappa(t'',t_2) = - \delta(t''-t_2) ,
\end{equation}
which has the `causal' solution

\begin{equation}
\kappa(t'',t_2) = - \Theta (t''-t_2)e^{-w(t''-t_2)}  ,
\label{kcaus}
\end{equation}
which leads to 

\begin{equation}
{\cal K}(t_2) = - \int_{t_2}^t dt'' e^{-w(t''-t_2)} \frac{\delta}{\delta \psi(t'')}
+ (z-1) e^{-w(t-t_2)} ,
\label{Ksol}
\end{equation}
as may be verified directly.

To evaluate

\begin{equation}
[\psi(t_1) \hat{\psi}(t_2)] = \int  \!{\cal D} \! \psi
\! \int \! {\cal D} \hat{\psi}  
\psi(t_1) {\cal K}(t_2) {\cal F}(\psi,\hat{\psi})~,
\label{brac2}
\end{equation}
we make use of our earlier
result, Eq. (\ref{delta}), to obtain:

\begin{eqnarray}
[\psi(t_1) \hat{\psi}(t_2)] &=& \int  \!{\cal D} \! \psi
\psi(t_1) \left\{
- \int_{t_2}^t dt'' e^{-w(t''-t_2)} \frac{\delta}{\delta \psi(t'')}
+ (z\!-\!1) e^{-w(t-t_2)} \right\}   \nonumber  \\
&\;\;\;\;\;\;\;\;\;\; \times& e^{\zeta +(z\!-\!1)\psi(t)}
\prod_{0<s \leq t} \delta(\psi(s)-\zeta e^{-ws}).
\label{brac3}
\end{eqnarray}
Now perform a functional integration by parts so that $\delta/\delta\psi(t'')$
operates on $ \psi(t_1) $.  This yields

\begin{eqnarray}
[\psi(t_1) \hat{\psi}(t_2)] \!\!\!\! &=& \!\! \int  \!{\cal D} \! \psi
\left[
\int_{t_2}^t dt'' e^{-w(t''-t_2)} \delta(t_1-t'') 
+ (z\!-\!1) e^{-w(t-t_2)} \psi(t_1) \right]
\nonumber \\
&\;\;\;\;\;\;\;\;\;\;\;\;\;\;\;\;\;\;\;\;\;\; \times &  
e^{\zeta +(z\!-\!1)\psi(t)}
\prod_{0<s \leq t} \delta(\psi(s)-\zeta e^{-ws}).
\label{brac4}
\end{eqnarray}
We may now integrate over $\psi$ to obtain

\begin{eqnarray}
[\psi(t_1) \hat{\psi}(t_2)] &=& \left[
(z\!-\!1)e^{-w(t-t_2)} \zeta e^{-wt_1} + \Theta(t_1-t_2)e^{-w(t_1-t_2)}
\right] U_t^0(z,\zeta) 
\nonumber \\
&\equiv& \langle \psi(t_1) \hat{\psi}(t_2) \rangle U_t^0(z,\zeta) \; ,
\label{brac5}
\end{eqnarray}
where
\begin{equation}
U_t^0(z,\zeta) \equiv e^{\zeta[1 + (z\!-\!1)e^{-wt}]}   .
\label{defu0}
\end{equation}
In light of the discussion following Eq. (\ref{ut}), we see that
in Eq. (\ref{defk}) and the subsequent expressions, $t_2$  should be
interpreted as $t_2^+$, which means that in the expectation,
Eq. (\ref{brac5}), we should take $\Theta(0) \equiv 0$.

To see what our results mean, consider the very simple situation of
${\cal L}_I = u\hat{\psi}\psi$, which is equivalent to exponential
decay with a rate $w'= w + u$.  Then,

\begin{equation}
e^{-S_I} = \exp\left[\!-\!u\int_0^tdt' \hat{\psi}\psi \right],
\end{equation}
and
\begin{eqnarray}
\nonumber 
U_t(z,\zeta) &=& U_t^0 \left( 1 
               - u\int_0^t dt' \langle \psi(t')\hat{\psi}(t')\rangle 
               + {\cal O}(u^2) \right)
\\
&=& U_t^0 \left[1 - u(z\!-\!1)\zeta t e^{-wt} 
+ {\cal O}(u^2) \right],
\label{ex1}
\end{eqnarray}
which leads to the exact result:

\begin{equation}
e^{\zeta[1 + (z\!-\!1)e^{-wt}e^{-ut}]}
= e^{\zeta[1 + (z\!-\!1)e^{-wt}]}[1 - ut\zeta(z\!-\!1)e^{-wt} + {\cal O}(u^2)].
\end{equation}
(Note that obtaining the correct result depends on
using $\Theta(0) = 0$.)

The preceding analysis exposes a curious feature of this formalism.
Normally in a Gaussian model, the expectation of the
field $\langle \psi \rangle$ is zero.  Here, by contrast, we have

\begin{equation}
[\psi(t_1)] = \int  \!{\cal D} \! {\psi}
\! \int \! {\cal D} \hat{\psi}  \psi(t_1) {\cal F} = \zeta e^{-wt_1}
U_t^0,
\label{exppsi}
\end{equation}
while from Eq. (\ref{brac3}) we have

\begin{equation}
[\hat{\psi}(t_1)] = \int  \!{\cal D} \! {\psi}
\! \int \! {\cal D} \hat{\psi}  {\cal K}(t_1) {\cal F} = (z\!-\!1)e^{-w(t-t_1)}
U_t^0.
\label{exppsih}
\end{equation}
This suggests that it would simplify matters if we were to introduce new
fields:

\begin{equation}
\varphi(\tau) = \psi(\tau) - \zeta e^{-w\tau}  ,
\end{equation}
and
\begin{equation}
\hat{\varphi}(\tau) = \hat{\psi}(\tau) - (z\!-\!1) e^{-w(t-\tau)},
\end{equation}
which by construction have expectation zero.
The boundary conditions on $\psi$ and $\hat{\psi}$ imply
that $\varphi(0) = \hat{\varphi}(t) = 0$.

Noting that $\partial_{t'} + w$ annihilates $e^{-wt'}$, we have

\begin{eqnarray}
-\int_0^t dt' \hat{\psi} (\partial_{t'} + w) \psi
&=& -\int_0^t dt' [\hat{\varphi} +(z\!-\!1)e^{-w(t-t')}]
(\partial_{t'} + w) \varphi
\nonumber \\
&=& -\int_0^t dt' \varphi (-\partial_{t'} + w) \hat{\varphi} \;
-(z\!-\!1) \varphi(t),
\end{eqnarray}
where in the second line we integrated by parts and used the fact that
$-\partial_{t'} + w$ annihilates $e^{wt'}$.  Integrating by parts once again
gives

\begin{equation}
-\int_0^t dt' \hat{\psi} (\partial_{t'} + w) \psi
= -\int_0^t dt' \hat{\varphi} (\partial_{t'} + w) \varphi
- (z\!-\!1)[\psi(t) - \zeta e^{-wt}] ,
\end{equation}
which, when inserted in Eq. (\ref{ut2b}), yields

\begin{eqnarray}
\nonumber
U_t(z,\zeta)\! &=& \!\! U_t^0(z,\zeta)
\int  \!{\cal D} \! \varphi
\! \int \! {\cal D} \hat{\varphi}
\exp \!\left[\! -\!\int_0^t\! \!dt'
\hat{\varphi}(\partial_{t'} \!+\! w)\varphi \right] e^{-S_I}
\\
& \equiv &  U_t^0(z,\zeta) \tilde{U}_t(z,\zeta)    \;.
\label{ut2c}
\end{eqnarray}
The change of variable, then, yields the unperturbed solution,
$U_t^0$, automatically, and eliminates
the boundary term from the exponential.  The normalization is such that
the functional integral is unity if $S_I = 0$.

We may now repeat the analysis of the basic contraction by letting

\begin{equation}
{\cal G}[\varphi,\hat{\varphi}] \equiv
\exp \!\left[\! -\!\int_0^t\! \!dt'
\hat{\varphi}(\partial_{t'} \!+\! w)\varphi \right],
\label{defg}
\end{equation}
and defining ${\cal J}$ such that

\begin{equation}
{\cal J}(t_2) {\cal G} = \hat{\varphi}(t_2) {\cal G} .
\label{defj}
\end{equation}
One readily verifies that

\begin{equation}
{\cal J}(t_2) = - \int_{t_2}^t dt'' e^{-w(t''-t_2)} 
\frac{\delta}{\delta \varphi(t'')},
\label{jt2}
\end{equation}
which leads directly to

\begin{eqnarray}
[\varphi(t_1) \hat{\varphi} (t_2)] & = &
U_t^0(z,\zeta) \Theta(t_1-t_2) e^{-w(t_1-t_2)}
\nonumber \\
& \equiv & U_t^0(z,\zeta) \; \langle \varphi(t_1) \hat{\varphi} (t_2) \rangle.
\label{conphi}
\end{eqnarray}
Then an average of $n$ fields (with equal numbers $\varphi$'s and
$\hat{\varphi}$'s) may be written

\begin{equation}
[\varphi(t_1) \hat{\varphi} (t_2) \cdots \varphi(t_{n-1}) \hat{\varphi} (t_n)]
= U_t^0(z,\zeta)\; \langle \varphi(t_1) \hat{\varphi} (t_2) \cdots
\varphi(t_{n-1}) \hat{\varphi} (t_n) \rangle,
\label{contn}
\end{equation}
with $ \langle \varphi(t_1) \hat{\varphi} (t_2) \cdots
\varphi(t_{n-1}) \hat{\varphi} (t_n) \rangle $ given by the sum of all
distinct products of pairwise contractions of the form of Eq. (\ref{conphi}).

Let us analyze the simple quadratic ``perturbation" using the ``$\varphi$"
representation.  Introducing the new fields we have

\begin{eqnarray}
-S_I &=& -\!u\int_0^tdt' \hat{\psi}\psi
\nonumber \\
     &=&-\!u\int_0^tdt' [\hat{\varphi} +(z-1)e^{-w(t-t')}]
        [\varphi +\zeta e^{-wt'}] ,
\end{eqnarray}
which gives us

\begin{equation}
U_t(z,\zeta) = U_t^{0'} \int  \!{\cal D} \! \varphi
\! \int \! {\cal D} \hat{\varphi} {\cal G}[\varphi,\hat{\varphi}]
\exp \!\left[\! -u\!\int_0^t\! \!dt'
\left(\hat{\varphi}\varphi +b \varphi +c\hat{\varphi} \right)
\right] ,
\label{uqpert}
\end{equation}
where 
\begin{eqnarray}
b(t') &=& (z-1)e^{-w(t-t')} ,
\\  \nonumber
\\
c(t') &=& \zeta e^{-wt'} ,
\end{eqnarray}
and
\begin{equation}
U_t^{0'} (z,\zeta) = \exp \left[ \zeta \left( 1+ 
(z-1)e^{-wt} (1-ut) \right) \right] \;.
\end{equation}

\noindent Let $U_t(z,\zeta) = U_t^{0'} (z,\zeta) \tilde{U} (z,\zeta)$.
The latter factor has the expansion:

\begin{eqnarray}
\tilde{U} (z,\zeta) &=& \sum_{n=0}^\infty (-\!u)^n
   \int_0^t dt_1 \int_0^{t_1} dt_2 \cdots \int_0^{t_{n-1}} dt_n
   \int  \!{\cal D} \! \varphi \! \int \! {\cal D} \hat{\varphi}
   \left[\hat{\varphi}_1 \varphi_1 +b_1 \varphi_1 +c_1 \hat{\varphi}_1 \right]
\nonumber \\
   &\;\;\;\; \times& 
   \left[\hat{\varphi}_2 \varphi_2 +b_2 \varphi_2 +c_2 \hat{\varphi}_2 \right]
   \cdots
   \left[\hat{\varphi}_n \varphi_n +b_n \varphi_n +c_n \hat{\varphi}_n \right]
   {\cal G}[\varphi,\hat{\varphi}] ,
\end{eqnarray}
where $\varphi_j \equiv \varphi(t_j)$, etc.  The $n=0$ term is unity (by
normalization), while the $n=1$ term vanishes, since
$\langle \varphi(t') \rangle = \langle \hat{\varphi}(t') \rangle 
= \langle \varphi(t') \hat{\varphi}(t') \rangle =0$.

For $n=2$, of the nine possible terms in ${\cal L}_I^2$, only one survives
the functional integral, giving

\begin{eqnarray}
u^2 \int_0^t dt_1 \int_0^{t_1} dt_2 b_1 c_2
\langle \varphi_1 \hat{\varphi}_2 \rangle &=& 
u^2 \zeta (z-1) \int_0^t dt_1 \int_0^{t_1} dt_2
e^{-w(t-t_1)} e^{-wt_2} e^{-w(t_1-t_2)}
\nonumber \\
   &=& \zeta (z-1) e^{-wt} \frac{(ut)^2}{2}
\end{eqnarray}
Next consider the $n=3$ contribution.  Once again there is but a single 
nonzero term, since we must choose $b_1 \varphi_1$ for the first factor 
(there is no $\varphi$ with time $>t_1$, with which to contract 
$\hat{\varphi}_1$), $c_3 \hat{\varphi}_3$ for the final factor (there is no 
$\hat{\varphi}$ with time $< t_3$, i.e., no way to contract $\varphi_3$), 
while the middle factor must be $ \hat{\varphi}_2 \varphi_2$.  Writing
this contribution in terms of functional derivatives we have

\begin{eqnarray}
-u^3 \int_0^t dt_1 \int_0^{t_1} d&t&_2 \int_0^{t_2} dt_3 b_1 c_3
   \int  \!{\cal D} \! \varphi \! \int \! {\cal D} \hat{\varphi}
\nonumber  \\
&\times&   \int_{t_2}^t dt'' e^{-w(t''-t_2)}  \int_{t_3}^t dt''' e^{-w(t'''-t_3)}
   \varphi_1 \varphi_2
  \frac{\delta}{\delta \varphi(t'')}  \frac{\delta}{\delta \varphi(t''')}
  {\cal G}[\varphi,\hat{\varphi}]
\label{u3}
\end{eqnarray}
Upon performing the two functional integrations by parts, 
$\delta/\delta \varphi(t'')$ may act on $\varphi_1$ and
$\delta/\delta \varphi(t''')$ on $\varphi_2$, or vice-versa.  In the 
former case we obtain the factor

\begin{equation}
\int_{t_2}^t dt'' e^{-w(t''-t_2)} \delta(t''-t_1) 
\int_{t_3}^t dt''' e^{-w(t'''-t_3)} \delta(t'''-t_2) = e^{-w(t_1-t_3)} .
\end{equation}
The $\delta$-functions can both be satisfied since  $t_1 \in [t_2,t]$
and $t_2 \in [t_3,t]$.  On the other hand, the second alternative
gives the factor

\begin{equation}
\int_{t_2}^t dt'' e^{-w(t''-t_2)} \delta(t''-t_2) 
\int_{t_3}^t dt''' e^{-w(t'''-t_3)} \delta(t'''-t_1) = 0 ,
\end{equation}
since $t'' = t_2$ does not fall within the range of integration.
Combining this result with the other factors the third-order term becomes
$\zeta (z-1) e^{-wt} (-ut)^3/3!$.

We now introduce a diagrammatic notation which will prove essential
in this and subsequent analyses.  The three terms in ${\cal L}_I$ are
represented by vertices, so:

\begin{eqnarray*}
b_i \varphi_i & \;\; \longleftrightarrow &  \;\; 
\stackrel{i}{\bullet} \!\!\!-\!\!\!\leftarrow 
\\  
c_i \hat{\varphi}_i \;\; & \longleftrightarrow & \;\;  
-\!\!\! \leftarrow \!\!\! \stackrel{i}{\bullet}  
\\  
\varphi_i \hat{\varphi}_i \;\; & \longleftrightarrow & 
\;\; -\!\!\! \leftarrow \!\!\! \stackrel{i}{\bullet} \!\!\!-\!\!\!\leftarrow 
\nonumber
\end{eqnarray*}
A contraction between $\varphi_i$ and $\hat{\varphi}_j$ (with $t_i > t_j$
due to the factor $\Theta(t_i -t_j)$) is represented by joining the
line exiting vertex $j$ with that entering vertex $i$.  Thus the $n=2$
and $n=3$ terms correspond to the diagrams:
\[
\stackrel{1}{\bullet} \!\!\!-\!\!\!\! \leftarrow \! \stackrel{2}{\bullet}
\;\;\;\;\; \mbox{and} \;\;\;\; 
\stackrel{1}{\bullet} \!\!\!-\!\!\! \leftarrow \! \stackrel{2}{\bullet}  
\!\!\!-\!\!\!\leftarrow \!\! \stackrel{3}{\bullet}  
\] 
respectively.  Associated with the line connecting vertices $i$ and $j$
(with $t_i > t_j$), is the factor $e^{-w(t_i-t_j)}$.

Each diagram constructed according to the following rules corresponds to
a term in the expansion of $\tilde{U}$.

\noindent 1) Draw $m \geq 1$ vertices of type
$\bullet \!\!\!-\!\!\!\leftarrow $ and an equal number of type
$-\!\!\! \leftarrow \!\!\! \bullet$.  Add $p \geq 0$ vertices of type
$-\!\!\! \leftarrow \!\!\! \bullet \!\!\!-\!\!\!\leftarrow$, for a total
of $2m+p$ vertices.

\noindent 2) Form all possible {\it unlabelled} diagrams, i.e., distinct
connections in which each outgoing line is joined with an ingoing line.

\noindent 3) Generate all distinct labellings of each unlabelled diagram
by assigning an index (time) to each vertex, such that the arrows always
point from the larger to the smaller index.

\noindent 4) A diagram consists of one or more connected parts; its 
contribution to $\tilde{U}$ is the product of factors
associated with these connected parts.  

In the example under consideration, the factor associated with
the $j$-th connected part (having $v_j$ vertices) is
\[
f_j = \zeta (z-1) e^{-wt} \frac{(-ut)^{v_j}}{v_j!}  \;.
\]

We now make use of a well known theorem in diagrammatic analysis
(see, e.g., \S 8.3 of Ref. \cite{binney}), which in the present context
states that $\ln \tilde{U}_t$ is given by the sum of all contributions
associated with {\it connected diagrams only}.
In the present case there is exactly one connected 
diagram at each order $n\geq2$, yielding

\begin{equation}
\ln \tilde{U}_t = \zeta (z-1) e^{-wt} \sum_{n=2}^\infty \frac{(-ut)^n}{n!} ,
\end{equation}
which, when inserted in Eq. (\ref{uqpert}), gives the exact result,

\begin{equation}
U_t (z,\zeta) = \exp \left[ \zeta \left( 1 + (z-1) e^{-(w+u)t} 
\right) \right] .
\label{uqpex}
\end{equation}

While the expansion of $\ln \tilde{U}_t $ in terms of connected graphs
represents a considerable simplification, it is possible, in principle,
to evaluate $\tilde{U}_t $ directly, without transforming to the
fields $\varphi$ and $\tilde{\varphi}$.  Consider, for example, the
second order term in the expansion of $\tilde{U}_t $, Eq. (\ref{ex1}):

\begin{eqnarray}
\nonumber
u^2 \int _0^t &\!& \!\!\!\!\!\! dt_1 \int_0^{t_1} dt_2 \int  \!{\cal D} \! \psi
\! \int \! {\cal D} \hat{\psi}
\hat{\psi}(t_1) \psi(t_1) 
\hat{\psi}(t_2) \psi(t_2) {\cal F}(\psi,\hat{\psi})
\\  \nonumber
&=& 
u^2 \int_0^t dt_1 \int_0^{t_1} dt_2 \int  \!{\cal D} \! \psi
\! \int \! {\cal D} \hat{\psi}
 {\cal F}(\psi,\hat{\psi})
\left[\int_{t_1}^t d\tau e^{-w(\tau-t_1)} \frac{\delta}{\delta \psi(\tau)} 
+ b(t_1) \right]
\\
&\;& \times
\left[\int_{t_2}^t d\tau' e^{-w(\tau'-t_1)} 
\frac{\delta}{\delta \psi(\tau')} + b(t_2) \right] 
\psi(t_1) \psi(t_2)  \;. 
\label{u2a}
\end{eqnarray}
Noting that the functional derivative w.r.t. $\psi(\tau)$ gives zero,
the above expression is seen to be

\begin{equation}
\frac{1}{2} u^2 t^2 \zeta (z\!-\!1) e^{-wt} \left[1 + 
\zeta (z\!-\!1) e^{-wt} \right] \;,
\label{u2b}
\end{equation}
which is precisely the ${\cal O}(u^2) $ contribution in the expansion
of $\tilde{U}_t$, Eq. (\ref{uqpex}).  Clearly, this analysis is 
considerably more cumbersome than the connected diagram expansion.

\section{Birth-and-death process}

Next we consider the more complex, but still exactly soluble problem of 
a birth-and-death process without saturation, i.e., Eq. (\ref{si}) with
$\nu = 0$.  Our first step, as before, is to rewrite the perturbation
in terms of the variables $\varphi$ and $\hat{\varphi}$:

\begin{eqnarray}
\nonumber
\lambda \int_0^{t} dt' \hat{\psi}^2 \psi
& = & \lambda \int_0^t dt' [\hat{\varphi} + b(t')]^2 [\varphi + c(t')]
\\ 
& = & \lambda \int_0^t dt' [\hat{\varphi}^2 \varphi+ 
2b \hat{\varphi} \varphi + b^2 \varphi + c\hat{\varphi}^2 
+ 2 b c \hat{\varphi} + b^2c]  \;.
\label{Lmu0}
\end{eqnarray}
In the second line, the final
term is independent of $\varphi$ and $\hat{\varphi}$.  Integrating it, and
combining it with the usual prefactor $U_t^0$, we have for this
problem

\begin{equation}
U_t^{0'} (z,\zeta) = \exp \left[\zeta \left( 1 + (z-1) e^{-wt}
  + \frac{\lambda}{w} (z-1)^2 e^{-wt} (1-e^{-wt}) \right) \right] \;.
\label{U0pmu0}
\end{equation}
The remaining terms will again be analyzed using diagrams.  There
are five vertices, as shown in Fig. 1 (note that the three we have encountered
before carry different factors here).

The first vertex in Fig. 1 (``terminal") must bear the lowest index  
of a given branch, since no lines exit from it.  The second and
fifth can only appear in the interior of a diagram, while the remaining two
can only appear as the $n$th of an $n$-vertex diagram, since
no lines enter.  Thus the expansion 
is not as complicated as it might at first appear.  The lowest-order diagram is
again $
\stackrel{1}{\bullet} \!\!\!-\!\!\!\! \leftarrow \! \stackrel{2}{\bullet}
$ which now corresponds to
\[
2 \lambda^2 (z-1)^3 \zeta  \int_0^t dt_1 \int_0^{t_1} dt_2 e^{-w(3t-2t_1-t_2)}
e^{-w(t_1-t_2)} e^{-wt_2}
= \left(\frac{\lambda}{w}\right)^2 \zeta (z-1)^3 e^{-wt} (1-e^{-wt})^2 \;.
\]

At third and higher orders there are graphs with bifurcations,
the simplest being the one shown in Fig. 2,
corresponding to
\begin{eqnarray}
\nonumber
2 \lambda^3 (z-1)^4 &\zeta&  \int_0^t dt_1 \int_0^{t_1} dt_2 \int_0^{t_2} dt_3
e^{-w(4t-2t_1-2t_2)}
e^{-w(t_1-t_3)} e^{-w(t_2-t_3)} e^{-wt_3}
\\ &=& \frac{2}{3!} \left(\frac{\lambda}{w}\right)^3 \zeta (z-1)^4 
e^{-wt} (1-e^{-wt})^3 \;.
\end{eqnarray}
The factor of 2 arises because there are two ways to contract the
$\hat{\varphi}$-lines exiting vertex 3.  
Thus all the vertices in this problem, except for the terminal $b^2 \varphi$,
carry a factor of 2, either explicitly, or due to the combinatorial
factor.
The other third-order connected diagram is
\begin{eqnarray}
\nonumber
4 &\zeta& (z-1)^4 \int_0^t dt_1 \int_0^{t_1} dt_2 \int_0^{t_2} dt_3 
e^{-w(4t-2t_1-t_2-t_3)}
e^{-w(t_1-t_3)} e^{-wt_3}
\\ 
&=& \frac{4}{3!} \left(\frac{\lambda}{w}\right)^3 \zeta (z-1)^4 
e^{-wt} (1-e^{-wt})^3 \;.
\end{eqnarray}
The two diagrams differ only by their numerical prefactors, the sum
of which just cancels the $1/3!$ that comes from the integrations.
We show in the Appendix that this
pattern continues at higher orders, i.e., that the sum of  
all numerical factors due to $n$-vertex diagrams is $n!$, so that

\begin{equation}
\ln \tilde{U}_t (z,\zeta) = \sum_{n=2}^\infty [(\lambda/w)(z-1)(1-e^{-wt})]^n \;.
\end{equation}
Combining this with Eq. (\ref{U0pmu0}), 
we obtain the exact result \cite{marro1},

\begin{equation}
U_t(z,\zeta) = \exp\left[\zeta \left( 1 + \frac{(z\!-\!1)e^{-wt}}
{1-(\lambda/w)(z\!-\!1)(1\!-\!e^{-wt})} \right) \right] \;.
\end{equation}

\section{Expansion of moments}

While the transformation to variables $\varphi$ and $\hat{\varphi}$,
with expectation zero, yields a certain simplifcation, it also
causes a proliferation in the number of vertices, so that in many
cases it may be advantageous to work with the original variables 
$\psi$ and $\hat{\psi}$.  We have seen that in this case the expansion
of $U_t$ becomes quite complicated.  Fortunately, it is still possible to
derive relatively simple expressions for the moments 
$\langle n^r (t) \rangle$ of the distribution.  

Note that from the definition of $\Phi_t(z)$ we have

\begin{equation}
\left. \langle n \rangle = 
\frac{\partial \Phi_t (z)}{\partial z} \right|_{z=1} \;,
\label{expn}
\end{equation}
and that, in general, the $r$-th {\it factorial moment}, 
$\langle n^r \rangle_f \equiv 
\langle n(n\!-\!1)(n\!-\!2) \cdots (n\!-\!r\!+\!1) \rangle $,
is given by:

\begin{equation}
\left. \langle n^r  \rangle_f =
\frac{\partial^r \Phi_t (z)}{\partial z^r} \right|_{z=1}\;.
\label{facmom}
\end{equation}
Using Eq. (\ref{map}) we then have

\begin{equation}
\langle n^r  \rangle_f = \int\frac{d \zeta d \zeta'}{2 \pi}
e^{-i \zeta \zeta'}
\left(\! \frac{\partial^r U_t (z,\zeta)}{\partial z^r} \! \right)_{z=1}
\Phi_0(i\zeta') \;.
\label{mapfm}
\end{equation}
If $U_t$ is of the form of Eq. (\ref{ut2b}), then

\begin{equation}
U_t^{(r)} (\zeta) \equiv \left(\!
\frac{\partial^r U_t (z,\zeta)}{\partial z^r} 
\! \right)_{z=1} =
e^\zeta  \int  \!{\cal D} \! {\psi}
\! \int \! {\cal D} \hat{\psi}
\psi(t)^r 
{\cal G}[\psi,\hat{\psi}] \exp[-S_I] \;.
\label{drUdzr}
\end{equation}
(Note that the prefactor $e^\zeta$ is just $U^0$ for $z \!=\! 1$.)
Thus the expectation of the variable $\psi$ is
closely related to that of $n(t)$ itself.

In case $n(0) = n_0$ (so that $\Phi_0(z) = z^{n_0}$), we have,
using Eq. (\ref{mapn0}),

\begin{equation}
\left.
\langle n^r (t) \rangle_f = \frac{\partial^{n_0} U_t^{(r)}}
 {\partial \zeta^{n_0}}
 \right|_{\zeta=0} \;.  
\label{fmn0}
\end{equation}
Evaluation of the above expression is facilitated by noting that
\[
\left. \frac{d^n}{d \zeta^n} \zeta^r e^\zeta \right|_{\zeta = 0}  
= \frac{n!}{(n\!-\!r)!}  \;.
\]  
If the initial distribution is Poisson with parameter $p$, then
Eq. (\ref{mappoi}) implies that 

\begin{equation}
\langle n^r (t) \rangle_f = e^{-p}
U_t^{(r)} (\zeta \!=\! p) \;.  
\label{fmpoi}
\end{equation}

It is instructive to evaluate the above expressions first for the
simple decay process, $S_I = 0$.  In this case

\begin{equation}
U_t^{(r)} (\zeta) =
[\psi(t)^r]_{z=1} = e^\zeta [\zeta e^{-wt}]^r \;,
\label{drUdzrdec}
\end{equation}
and a short calculation shows that for the initial condition
$n(0) = n_0$,

\begin{equation}
\langle n (t) \rangle = n_0 e^{-wt}   \;,
\end{equation}
and
\begin{equation}
\mbox{ Var}[n (t)] = n_0 e^{-wt} (1 - e^{-wt})  \;,
\end{equation}
as expected.  For the Poisson initial distribution, one finds that

\begin{equation}
\langle n (t) \rangle = \mbox{Var}[n (t)] = p e^{-wt}   \;,
\end{equation}
consistent with the probability distribution $p_n(t)$ 
being Poisson with parameter $p e^{-wt}$.

Next we consider the moments for the birth-and-death process.
In this case 
\[
-S_I = \lambda \int_0^t dt' \hat{\psi}^2 \psi
\]
corresponding to a vertex that branches to the left.  To evaluate
$U_t^{(r)} (\zeta)$, we expand
$e^{-S_I}$, generating diagrams consisting
of vertices and a single ``sink", with $r$ lines entering,
to the left of all vertices.
Each line entering a node (i.e., a vertex or the sink), corresponds 
to a variable $\psi (t')$; if not contracted, it contributes a factor
of $\zeta e^{-wt'}$.  (We call such uncontracted lines ``external".
Recall that for the birth-and-death process,
$w \equiv \mu \!=\! \lambda $.)  On the other hand, lines 
{\it exiting} a vertex correspond to $\hat{\psi}$ variables, and
{\it must} be contracted with a $\psi$ line, because
$[\hat{\psi}] = 0$ when $z \!=\! 1$.  With $z \!=\! 1$,
the basic contraction is:

\[
\langle \psi(t_1) \hat{\psi}(t_2) \rangle =
\Theta(t_1-t_2)e^{-w(t_1-t_2)}    \;.
\] 
Thus the perturbation series for a given moment is much simpler
than the full series for $U_t (z,\zeta)$.

Consider the case $r \!=\! 1$.  The sink has only a single line
entering, so diagrams with $n \geq 1$ vertices all give zero, and
the only nonzero contribution to 
$(\partial U_t /\partial z)_{z=1}$
is for $n \!=\!0$ (the sink itself),
which is $\zeta e^\zeta e^{-wt}$.

For $r \!=\! 2$ the sink has two lines.  There is then the $n \!=\! 0$
contribution, $\zeta^2 e^\zeta e^{-2wt}$, and the 1-vertex diagram
shown in Fig. 3.  Including the combinatorial factor of 2, its
contribution is

\[
2 \lambda e^\zeta \int_0 ^t dt_1 
\langle \psi (t) \hat{\psi} (t_1) \rangle^2 \zeta e^{-wt_1}
= \frac{2 \lambda}{w} \zeta e^\zeta e^{-wt}(1 - e^{-wt})
\]
Using these results, we readily find that 
\begin{equation}
\langle n(t) \rangle = \langle n(0) \rangle e^{-wt} \;.
\end{equation}
For $n(0) = n_0$, the variance is
\begin{equation}
\mbox{Var}[n(t)] = n_0 \frac{\mu \!+\! \lambda }{w} e^{-wt}(1 - e^{-wt}) 
\end{equation}
for $w \neq 0$, while for $w \!=\! 0$ we have
\begin{equation}
\mbox{Var}[n(t)] = 2 \lambda n_0 t \;.
\end{equation}
For a Poisson initial distribution,
\begin{equation}
\mbox{Var}[n(t)] = p e^{-wt} \left[ 1 + 
\frac{2 \lambda }{w} (1 - e^{-wt})  \right]   \;,
\end{equation}
which becomes $p(1 \!+\! 2 \lambda t)$ in case $w \!=\! 0$.
We see that at long times, the variance decays (grows) exponentially, 
for $w > 0$ ($w < 0$), and that for $ w \!=\! 0$ the variance
grows $ \sim t$, reflecting the diffusive character of the process
in this case.  Higher-order moments may be evaluated similarly;
only diagrams with $ n \leq r \!-\! 1 $ vertices contribute to
$\langle n^r  \rangle_f$.

\section{Malthus-Verhulst Process}

We now return to the Malthus-Verhulst
process; the three vertices associated with $-S_I$ [see Eq. (\ref{si})]
are shown in Fig. 4.  The series for  
$U_t^{(r)} (\zeta) $ 
begins with the ``sink" term, $\zeta^r e^\zeta e^{-rwt}$,
and then includes diagrams with
$n \geq 1$ vertices.  (Note that the $n$-th order contributions
carry factors of $(-\nu)^s \lambda^{n-s}$, with $s$ variable.)
Based on our diagrammatic analysis of the
birth-and-death process, we can formulate the following simple rules
for evaluating the $n$-th order contribution to $U_t^{(r)}$:                                         

\noindent 1) Draw all diagrams consisting of $n$ vertices and
a single $r$-line sink to the left of all vertices.  Each line
exiting a vertex must be contracted with a line entering another vertex
to the left.

\noindent 2) The sink has time variable $t$; assign time variables
$t_1$,...,$t_n$ to the vertices from left (nearest the sink) to right.

\noindent 3) For each external line include a factor of
$\zeta e^{-w\tau}$, where $\tau$ is the time variable of the associated
vertex (if the external line is attached to the sink, $\tau \!=\! t$).
For each internal line from vertex $j$ to vertex $i$, ($i<j$) 
include a factor of $e^{-w(t_i - t_j)}$.

\noindent 4) Include the factors ($\lambda$ or $-\nu$)
associated with each vertex, and the combinatorial factor associated with
the number of ways of realizing the contractions.  Finally, integrate
over the time variables $t_1$,...,$t_n$, with
\[
t \geq t_1 \geq t_2 \geq \cdots \geq t_n \geq 0  \;.
\]
(Note that by fixing the time ordering we effectively include the
factor $1/n!$ that comes from expanding $e^{-S_I}$.)

Let us consider some examples in the evaluation of $\langle n \rangle$;
the low-order diagrams are shown in Fig. 5.  For $n\!=\!1$ we
have only diagram (a), whose contribution is:
\[
-\nu \zeta^2 e^\zeta \int_0^t dt_1 e^{-w(t-t_1)} e^{-2wt_1}
= -\frac{\nu}{w} \zeta^2 e^\zeta e^{-wt} (1 \!-\! e^{-wt})  \;.
\]
At second order we have the following contributions.  Diagram (b):

\begin{eqnarray*}
-2 &\!& \!\!\!\!\!\! \lambda \nu \zeta e^\zeta \int_0^t dt_1 \int_0^{t_1} dt_2
e^{-w(t-t_1)} e^{-2w(t_1-t_2)} e^{-wt_2}
\\
&=& - \frac{2\lambda \nu}{w^2} \zeta e^\zeta e^{-wt}
\left[ wt - 1 + e^{-wt} \right].
\end{eqnarray*}
Diagram (c):

\begin{eqnarray*}
2 &\!& \!\!\!\!\!\! \nu^2 \zeta^2 e^\zeta \int_0^t dt_1 \int_0^{t_1} dt_2
e^{-w(t-t_1)} e^{-2w(t_1-t_2)} e^{-2wt_2}
\\
&=& - \frac{2 \nu^2}{w^2} \zeta^2 e^\zeta e^{-wt}
\left[ wt - 1 + e^{-wt} \right].
\end{eqnarray*}
Diagram (d):

\begin{eqnarray*}
2 &\!&  \!\!\!\!\!\! \nu^2 \zeta^3 e^\zeta \int_0^t dt_1 \int_0^{t_1} dt_2
e^{-w(t-t_1)} e^{-w(t_1-t_2)} e^{-wt_1} e^{-2wt_2}
\\
&=& \frac{\nu^2}{w^2} \zeta^3 e^\zeta e^{-wt}
\left[1 - e^{-wt} \right]^2 .
\end{eqnarray*}
For fixed $n(0)$ this yields the expansion:

\begin{eqnarray}
\nonumber
\langle n(t) \rangle &=& n_0 e^{-wt} \left[ 1 - 
\frac{\nu}{w} (n_0 \!-\!1) (1 \!-\! e^{-wt})   \right.
\\
&\;\;\;\; -& \left. \frac{\nu}{w^2} \left( 2[\lambda + (n_0 \!-\! 1) \nu]
[e^{-wt} \!-\! 1 \!+\! wt] - \nu (n_0 \!-\! 1)(n_0 \!-\! 2)
(1 \!-\! e^{-wt})^2 \right)+ \cdots \right] .
\label{denmv}
\end{eqnarray}
For $r\!=\!2$ a similar calculation results in

\begin{eqnarray}
\nonumber
\langle n(t)[n(t) \!-\! 1] \rangle &=& n_0 (n_0 \!-\!1) e^{-2wt} 
 - 2\nu  n_0 (n_0 \!-\!1) t e^{-2wt}
\\
&\;\;\;\; +& \frac{2}{w^2} n_0 e^{-wt} (1 \!-\! e^{-wt})
\left[ \lambda - \nu (n_0 \!-\! 1)(n_0 \!-\! 2) e^{-wt} \right] 
+ \cdots .
\label{fm2mv}
\end{eqnarray}

The evaluation of diagrams can be simplified somewhat by considering
the Laplace transform of $\langle n^r \rangle_f$:  
\[
\langle n^r (s) \rangle_f = \int_0^\infty dt e^{-st} 
\langle n^r (t) \rangle_f
\]
We have
seen that each line entering a node with time variable $\tau$
carries a factor of $e^{-w\tau}$, and each line exiting the
node carries a factor of $e^{w\tau}$.  Letting $\ell_i$ be
the number of lines entering node $i$ less the number exiting,
we can write the time integration factor for the general diagram so:
\[
e^{-rwt} \int_0^t dt_1 \cdots \int_0^{t_{n-1}} dt_n
\exp [ -w(\ell_1 t_1 + \cdots + \ell_n t_n)] .
\]
Inverting the order of the integrations, the Laplace transform of
the above expression becomes:
\[
\int_0^\infty dt_n \int_{t_n}^\infty dt_{n-1} \cdots \int_{t_2}^\infty dt_1
\int_{t_1}^\infty dt e^{-(s+rw)t}
\exp [ -w(\ell_1 t_1 + \cdots + \ell_n t_n)]
\]
\[
= \frac{1}{s \!+\!rw} \frac{1}{s \!+\!(r\!+\!\ell_1)w}  \cdots
\frac{1}{s \!+\!(r\!+\!\ell_1+\cdots \!+\! \ell_n)w}   .
\]

\subsection{A numerical example}

The diagramatic expansion for the Malthus-Verhulst process can be
extended, and simplified (by identifying the so-called {\it reducible}
diagrams \cite{binney}), but such analysis is beyond the scope of
this article and will be defered to a future work.  We close this
section with an application of the second-order expansion for
the mean population, Eq. (\ref{denmv}).  Note that for short times,
the factor multiplying $n_0 e^{-wt}$ may be written as
\[
f(t) = 1 - At + Bt^2 + {\cal O}(t^3)    ,
\]
with $A = \nu (n_0 \!-\!1)$ and
$B = \nu [(n_0 \!-\!1)(n_0 \!-\!3)\nu - \lambda]$    .
Evidently, our analysis generates an expansion in powers of $t$,
whose convergence would seem to require that $\nu n_0 t \ll 1$.

If we wish to extend the range of validity of the expansion, we must
transform to a new variable that remains finite as $t \to \infty$;
a glance at Eq. (\ref{denmv}) suggests that we use
$ y \!=\! 1 \!-\! e^{-wt}$ as the new variable. 
(This transformation is very useful in analysis of 
series, for example in the study of 
random sequential adsorption \cite{rsa}.)  
In fact, $f(t)$ is readily expressed in terms of $y$:
\begin{equation}
f(t) = 1 - \overline{A} y + \overline{B} y^2 + {\cal O}(y^3)
\label{fy}
\end{equation}
with $\overline{A} = A/w$ and $\overline{B} = B/w^2$
To proceed, we form a Pad\'e approximant \cite{baker,guttmann} 
to the power series in $y$:
\begin{equation}
f(y) = \frac{1 + ay}{1+by} .
\label{pade}
\end{equation}
Equating coefficients of $y$ and of $y^2$, we find
\[
b = \frac{\overline{B}}{\overline{A}} 
= \frac{\nu (n_0\!-\!1) (n_0\!-\!3) - \lambda}{w (n_0\!-\!1)}
\]
and
\[
a = b - \overline{A}   
= - \frac{2\nu (n_0\!-\!1) + \lambda}{w (n_0\!-\!1)}.
\]
(Here we must note that a Pad\'e approximant to a series of
three terms can only serve as a very rough approximation!)

We apply this expression to the Malthus-Verhulst process with parameters 
$\mu = 1$, $\lambda = 0.5$, and $\nu = 0.1$.  
In Fig. 6 we compare the numerical solution of the
master equation with the perturbation theory expression,
$\langle n(t) \rangle = n_0 e^{-wt} f(t)$ (with $f$ represented
by the Pad\'e approximant), and with the simple exponential decay,
$n_0 e^{-wt}$, for $n_0 = 3$ and $n_0 = 10$. 
The perturbation expansion,
using the Pad\'e expression, yields a good approximation to the
correct value, despite the very short series used, and the fact
that for $n_0 = 10$ we have $\nu n_0 = 1$, so that the time series
has only a small radius of convergence.

\section{Coupled Malthus-Verhulst processes}

In this section we show how the path integral formalism can
be applied to processes on a lattice, leading to a field theory
in the continuum limit.  The process of interest is a lattice
of coupled Malthus-Verhulst processes, with diffusive exchange
of particles between neighboring cells.  
This process describes the dynamics of a population distributed in
space, with the individuals performing random walks, allowing the
population to spread.  In the continuum limit, the process is
of great interest as a field theory for directed percolation
\cite{cardy96}.

At each site {\bf r} of the lattice,
there is a process $n_{\bf r}(t)$ whose evolution is given by
Eq. (\ref{opmv}).  The new feature in this model is diffusion,
represented by the operator:

\begin{equation}
L_D = D \sum_{\bf r} \sum_{\bf e} (\pi_{\bf r+ e} - \pi_{\bf r}) a_{\bf r}.
\label{diffop}
\end{equation}
where $D$ is the diffusion rate and $\sum_{\bf e}$ is over the
vectors from a given site to its nearest neighbors.

The evolution kernel $U_t$ is a function of the variables $z_{\bf r}$
and $\zeta_{\bf r}$ defined at each site; we use $\{z\}$ and $\{\zeta\}$
to denote these sets of variables.
To write $U_t$ for this process we add the diffusive contribution
to the Malthus-Verhulst part, Eq. (\ref{ut2}), to find

\begin{eqnarray}
\nonumber
U_t(\{z\},\{\zeta\})\! &=& \!\! \int  \!{\cal D} \! {\psi}
\! \int \! {\cal D} \overline{\psi}
\! \exp \!\left[\! -\!\int_0^t\! \!dt'\sum_{\bf r}
 \left( \overline{\psi}_{\bf r} \dot{ \psi}_{\bf r} 
 -D\overline{\psi}_{\bf r} \Delta \psi_{\bf r} 
\right. \right.
\\
&-& \left. \left. 
\lambda (\overline{\psi}_{\bf r}\!-\!1)\overline{\psi}_{\bf r} \psi_{\bf r}\! \!
- \! \mu(1\!-\!\overline{\psi}_{\bf r})\psi_{\bf r} 
\!-\! \nu (1\!-\!\overline{\psi}_{\bf r})\overline{\psi}_{\bf r} \psi_{\bf r}^2
 \right)\! \!+ \sum_{\bf r} z_{\bf r} \psi_{\bf r}(t) \right] ,
\label{utmvl}
\end{eqnarray}
where $\Delta \psi_r = \sum_{\bf e} (\psi_{\bf r+e} - \psi_{\bf r})$
is the discrete Laplacian, and the functional integrals are now understood
to include the funcions $\psi_{\bf r}$ and $\overline{\psi}_{\bf r}$ at each
site.
Eliminating the linear term as usual, by letting
$\hat{\psi} = i\psi' -1$, and introducing $w \!=\! \mu \!-\! \lambda$, 
we obtain

\begin{eqnarray}
\nonumber
U_t(\{z\},\{\zeta\})\! &=& \!\! \int  \!{\cal D} \! {\psi}
\! \int \! {\cal D} \hat{\psi}
\! \exp \!\left[\! -\!\int_0^t\! \!dt'\sum_{\bf r}
 \left( \hat{\psi}_{\bf r}( \partial_{t'} + w - D\Delta){ \psi}_{\bf r} 
\right. \right.
\\
&-& \left. \left. 
\lambda \hat{\psi}_{\bf r}^2 \psi_{\bf r}
\!+\! \nu \hat{\psi}_{\bf r}(1 \!+\! \hat{\psi}_{\bf r}) \psi_{\bf r}^2
 \right)\! \!+ \sum_{\bf r}[\zeta_{\bf r} \!+\! 
 (z_{\bf r} \!-\! 1) \psi_{\bf r}(t)] \right] .
\label{utmvl2}
\end{eqnarray}

Up to this point, our expression for $U_t$ is exact.  We now make
a number of simplifications, leading to an effective action for the
lattice of coupled Malthus-Verhulst processes.  
(Each deserves a careful justification,
but we shall not enter into such questions here.)

\noindent i) We drop the boundary term, which should not influence
stationary properties.

\noindent ii) We take the continuum limit, so that $\psi_{\bf r}(t)
\to \psi({\bf x}, t)$, and $\Delta \psi_{\bf r} \to \nabla^2 \psi$.

\noindent iii) We discard the term $\propto \hat{\psi}^2 \psi^2$,
which turns out to be {\it irrelevant} to the scaling behavior
near the critical point \cite{janssen}.

Under these approximations the argument of the exponential 
in Eq. (\ref{utmvl2}) becomes the effective action

\begin{equation}
S = \int dt' \int d^d x \left\{ \hat{\psi}(\partial_{t'} + w -D \nabla^2)
\psi + \nu \hat{\psi} \psi^2 - \lambda \psi \hat{\psi}^2 \right\} .
\label{actdp}
\end{equation}
This is the action corresponding to directed percolation \cite{cardy96}
or Reggeon field theory (a particle physics model with the same
formal structure as the continuum theory of spatially extended
population) \cite{cardy,grassberger}.   
This theory has been analyzed using renormalization
group techniques, to show that the upper critical dimension $d_c \!=\!4$,
and to derive expressions for critical exponents in an expansion
in $\epsilon \!=\! 4 \!-\! d$ \cite{janssen,wijland}.  While such
developments lie beyond the scope of this article, we can get a
qualitative understanding of the physics represented by $S$ by 
ignoring, for the moment, the term $\propto \psi \hat{\psi}^2 $.
Functional integration over $\hat{\psi}$ then imposes the following
partial differential equation as a constraint on $\psi$:  

\begin{equation}
\frac{\partial \psi}{\partial t} = D \nabla^2 \psi - w \psi - \nu \psi^2 ,
\label{dpmft}
\end{equation}
which is, in fact, the mean-field theory of directed percolation
and allied models \cite{marro5}.  We see that this equation
correctly predicts an {\it absorbing state}, $\psi = 0$.  For
$w \geq 0$ the solution flows to this state, regardless of the
initial condition.  For $w < 0$, however, another stationary state
appears: the uniform solution $\psi = |w|/\nu$.  Thus $w \!=\! 0$
marks a continuous phase transition.   The field $\psi ({\bf x},t)$
may be interpreted as a local population density.
This mean-field description is readily shown
to yield the critical exponents $\beta \!=\! 1$ (for the order
parameter $\langle \psi \rangle$), $\nu_{||} \!=\! 1$ (for
the relaxation time), and $\nu_\perp = 1/2$
(for the correlation length) \cite{marro5}.

In this case, the effect of the neglected term in the action, 
$- \lambda \psi \hat{\psi}^2 $, can be represented
by a noise term $\eta({\bf x},t) $ in Eq. (\ref{dpmft}), which
now becomes a Langevin equation or stochastic partial differential
equation:
\begin{equation}
\frac{\partial \psi}{\partial t} = D \nabla^2 \psi - w \psi - \nu \psi^2 
+ \eta({\bf x},t)   ,
\label{dpspde}
\end{equation}
where $\eta$ is a Gaussian noise with $\langle \eta({\bf x},t) \rangle = 0$
and autocorrelation proportional to the local density \cite{cardy96}:

\begin{equation}
\langle \eta({\bf x},t) \eta({\bf y},s)
= \lambda \psi({\bf x},t) \delta^d ({\bf x} - {\bf y}) \delta(t-s)   ,
\label{autocorr}
\end{equation}
(note that in this way the noise respects the absorbing state).
In the presence of noise, the critical point is renormalized from
its mean-field value of $w_c = 0$, and, more significantly,
the critical exponents take non-mean-field values for $d < 4$.

As we have noted, the transcription of a stochastic model to
a Langevin equation is not always straightforward.  For this reason,
the path-integral formalism discussed in this article is especially
valuable: it allows on to construct an action (that is, the
starting point for a renormalization group analysis), without
having to postulate noise properties.   In addition to the
analysis of directed percolation discussed above, recent applications
of the method include the annihilation reactions $kA \to 0$
\cite{peliti86,lee}, and
branching and annihilating random walks \cite{baw}.

\section{Summary}

We have reviewed the formalism, based on the work of Peliti, that
maps a Markov process to a path integral representation, and
have presented detailed examples of its application to birth-and-death
processes.  We show, in particular, how the exact solutions
for the decay and simple birth-and-death process can be recovered,
and derive a perturbation expansion for moments in the Malthus-Verhulst
process.  Finally, we show how the evolution kernel for a lattice of
coupled Malthus-Verhulst processes leads, in the continuum limit, to
a field theory for directed percolation.  As the study of nonequilibrium
processes grows, we expect the methods discussed here
to attract ever-greater interest.
\vspace{1em}

\noindent{\bf Acknowledgments}

We are grateful to Migual A. Mu\~noz and Ad\'elcio Carlos de Oliveira 
for helpful comments. This work was supported by CNPq.
\vspace{1em}

{\large {\bf Appendix}}
\vspace{1em}

We aim to show that the sum of all numerical weights
associated with $n$-vertex diagrams in the expansion of
$\ln \tilde{U}_t $ for the birth-and-death process is $n!$.
To see this, consider
an arbitrary $n,b$-diagram, i.e., one having $n$ vertices, 
$b$ of them bifurcations.
There is a single factor of $\zeta $, associated with vertex $n$.
Such a diagram will have $b+1$ terminal vertices (carrying factors $b_i^2$), 
and so $n-2b-1$ nonterminal vertices that are not bifurcations, which
carry factors $b_j$.  Each $b_k$ carries a factor of $e^{-wt} (z-1)$ 
for a total of $n+1$ such.  Next consider the factors
$e^{wt_i}$ for each vertex $i$.  If $i$ is terminal, there are two
such factors (from $b_i^2$), and a factor $e^{-wt_i}$ due to the line
entering the vertex.  If $i<n$ is neither terminal nor a bifurcation,
there is a factor $e^{wt_i}$ from $b_i$, while the exponential factors
associated with the lines entering and leaving the vertex
cancel.  If $i<n$ is a bifurcation, there is no factor $b_i$, but
there is again a net factor of $e^{wt_i}$ as two lines exit, while
only one enters vertex $i$.  Consider vertex $n$.  If it
is a bifurcation, then the net factor is $e^{wt_n}$ (two lines 
exiting, factor $e^{-wt_n}$ in $c_n$).  The same holds 
if vertex $n$ is not a bifurcation (single line exiting, factors
in $b_i$ and $c_i$ cancel).  In summary, there is a factor $e^{wt_i}$
associated with {\it each} vertex.  Finally, all vertices carry a
factor of 2, except for the terminal ones, leading to an overall
factor of $2^{n-b-1}$.  Combining all of these observations, the
contribution due to a given labelled $n,b$-diagram is
\begin{eqnarray}
\nonumber
2^{n-b-1} &\zeta& (z-1)[\lambda (z-1)]^n e^{-(n+1)t} 
\int_0^t dt_1 \cdots \int_0^{t_{n-1}} dt_n 
e^{w(t_1+\cdots + t_n)}
\\ &=& \frac{2^{n-b-1}}{n!} \left(\frac{\lambda}{w}\right)^n \zeta (z-1)^{n+1} 
e^{-wt} (1-e^{-wt})^n \;.
\end{eqnarray}

To find the contribution $\propto \lambda^n$, it remains to evaluate
the sum of all such terms, that is, to determine
\begin{equation}
W(n) = \sum_{G_n} 2^{n-b-1}
\label{wn}
\end{equation}
where the sum is over all distinct labelled diagrams of $n$ vertices.
A labelled $n,b$-diagram has a {\it weight} of $ 2^{n-b-1}$.
Let $W(n,b)$ be the sum the weights of all $n,b$-diagrams. 
Define the {\it degree} of a vertex as the number
of lines that exit from it, so that a terminal vertex has degree 0,
bifurcations have degree 2, and all other vertices have degree 1.
(In this model there are no vertices with degree $>2$.)

Now, given a labelled $n,b$-diagram,
we can generate a set of distinct $n+1$-vertex labelled diagrams by
the following recipe:

\noindent i) Relabel the vertices 1,...,n as 2,...,n+1.

\noindent ii) Attach a new vertex (`1') to any
vertex of degree less than 2.  

\noindent It is easy to see that (1) each choice for attaching the new vertex generates
a different diagram; (2) the sets generated by different $n$-vertex labelled 
diagrams are mutually disjoint;
(3) applied to the complete set of $n$-vertex labelled diagrams, the
procedure generates the complete set of $n\!+\!1$-vertex diagrams.
When we attach the new vertex to one of degree 0, the number of bifurcations
does not change, so the new diagram has an additional factor of 2 in its
weight.  If on the other hand we attatch the new vertex to one of degree 1,
we generate a new bifurcation, and the weight remains unchanged.   
Recalling that an $n,b$-diagram has $b\!+\!1$ vertices of degree zero, and 
$n\!-\!2b\!-\!1$ of degree 1, we 
have for $n > 2$ the following recurrence relation:

\begin{equation}
W(n,b) = 2(1\!+\!b)W(n\!-\!1,b) + (n\!-\!2b)W(n\!-\!1,b\!-\!1) \;.
\label{recrel}
\end{equation}
Starting with $W(2,0) = 2$ (and $W(2,j) =0$ for $j>0$), we readily find
$W(3,0) = 4$, $W(3,1) = 2$, and then $W(4,0) = 8$, $W(4,1) = 16$, and so
on.  

To solve for $W(n)$ in general, we introduce a generating function
\begin{equation}
g(x,y) = \sum_n \sum_b x^n y^b \frac{W(n,b)}{n!} \;.
\label{diagen}
\end{equation}
(Since $W(n)$ appears to grow like $n!$ we need the factorial for $g$
to have a nonzero radius of convergence.  Without it, we get a function
with an essential singularity at $x\!=\!0$, as the reader may verify!)   
For purposes of
analysis, it is convenient to define $W(1,0) \equiv 1$ and $W(1,j) \equiv 0$
for $j>0$, which is completely consistent with the  
recurrence relation for $n=2$.  (There is, of course, no diagram with $n=1$.)
For $n=0$, $W$ vanishes identically, so we must add a source term
in Eq. (\ref{recrel}).  The 
modified recurrence relation, valid for $n=1, 2, 3,...$ and $b \geq 0$, is

\begin{equation}
W(n,b) = 2(1\!+\!b)W(n\!-\!1,b) + (n\!-\!2b)W(n\!-\!1,b\!-\!1) 
+ \delta_{n,1} \delta_{b,0} \;.
\label{mrecrel}
\end{equation}
Multiply this relation by $x^n y^b/n!$ and sum over $n$ and $b$.  Letting
$m = n\!-\!1$ and rearranging, one finds

\begin{eqnarray}
\nonumber
g(x,y) &=& 2 \sum_{m \geq 0} \sum_{b \geq 0} \frac{x^{m+1} y^b}{(m\!+\!1)!} 
\left[(1\!+\!b)W(m,b) -b W(m,b\!-\!1) \right] 
\\ &\;\;\; +&  x \sum_{m \geq 0} \sum_{b \geq 0} \frac{x^m y^b}{m!} W(m,b\!-\!1)
       + x \;.
\label{rr1}
\end{eqnarray}
Now let $b'=b\!-\!1$ in the terms multiplying $W(m,b\!-\!1)$.  After a
simple rearrangement (and dropping the primes) we have

\begin{equation}
g(x,y) = 2 (1-y)\left(1+ y\frac{\partial}{\partial y}\right)
\sum_{n \geq 0} \sum_{b \geq 0} \frac{x^{n+1} y^b}{(n\!+\!1)!} W(n,b) 
       +  xy g(x,y) + x \;.
\label{rr2}
\end{equation}
We require 

\begin{equation}
G(x) \equiv \sum_{n \geq 1} \frac {x^n W(n)}{n!} = g(x,1) \;.
\label{bigG}
\end{equation}
For $y=1$, Eq. (\ref{rr2}) immediately yields $G(x) = x/(1\!-\!x)$,
implying $W(n) = n!$.

\newpage

\noindent FIGURE CAPTIONS
\vspace{1em}

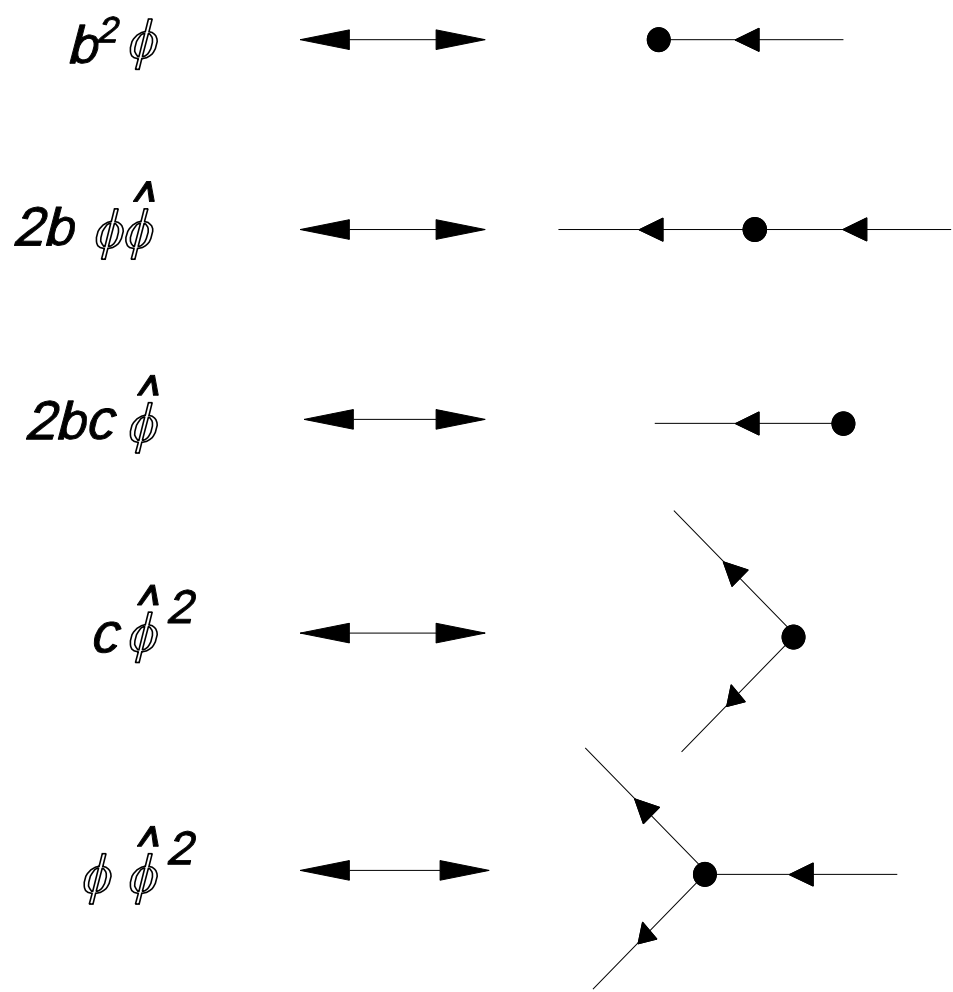
\noindent FIG. 1. Vertices in the birth-and-death process.
\vspace{1em}

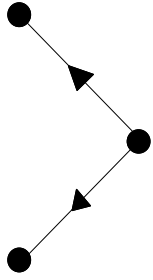
\noindent FIG. 2. An ${\cal O}(\lambda^3)$ diagram in the
birth-and-death process.
\vspace{1em}

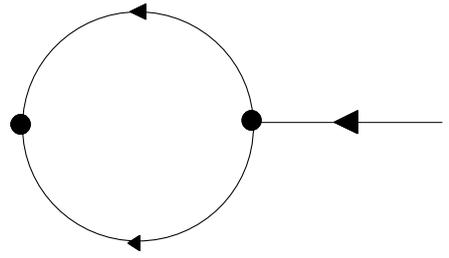
\noindent FIG. 3. One-loop diagram in the expansion of
$\langle n^2 \rangle_f$ for the birth-and-death process.
\vspace{1em}

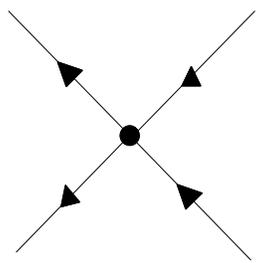
\noindent FIG. 4. Vertices in the moment expansion for the 
Malthus-Verhulst process.
\vspace{1em}

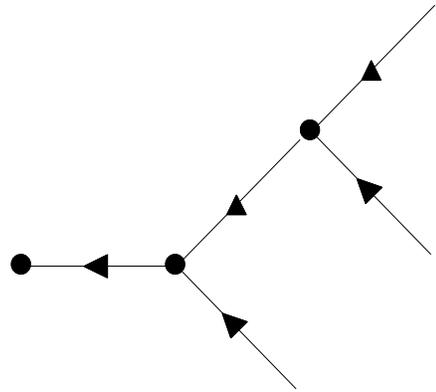
\noindent FIG. 5. Low-order diagrams in the expansion of
$\langle n \rangle$ for the Malthus-Verhulst process.
\vspace{1em}

\noindent FIG. 6. Mean population size in the Malthus-Verhulst
process with $\mu = 1$, $\lambda = 0.5$, $\nu = 0.1$;
$n_0 \!=\! 3$ (upper); $n_0 \!=\! 10$ (lower).
In each panel, the middle curve represents the exact 
(numerical) solution, the lower, the Pad\'e approximant to the
second-order perturbation series, and the upper, simple
exponential decay, $\langle n \rangle = n_0e^{-wt}$.
\vspace{1em}

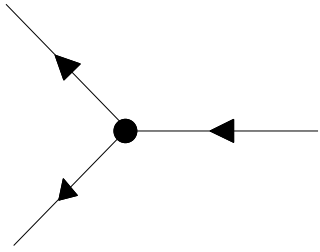
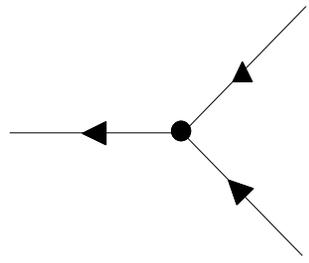
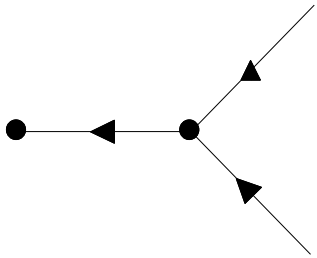
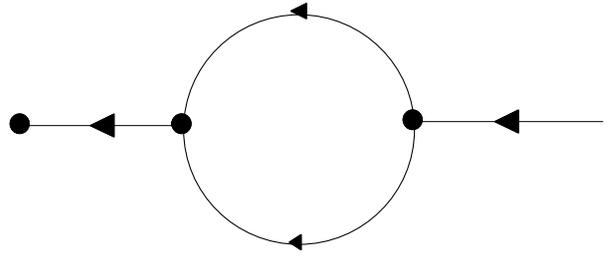
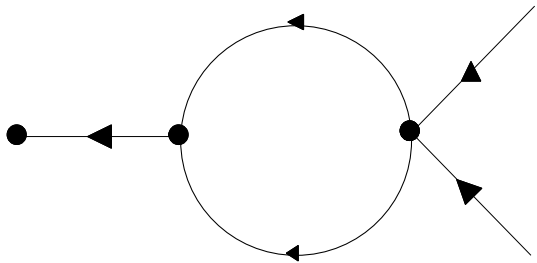
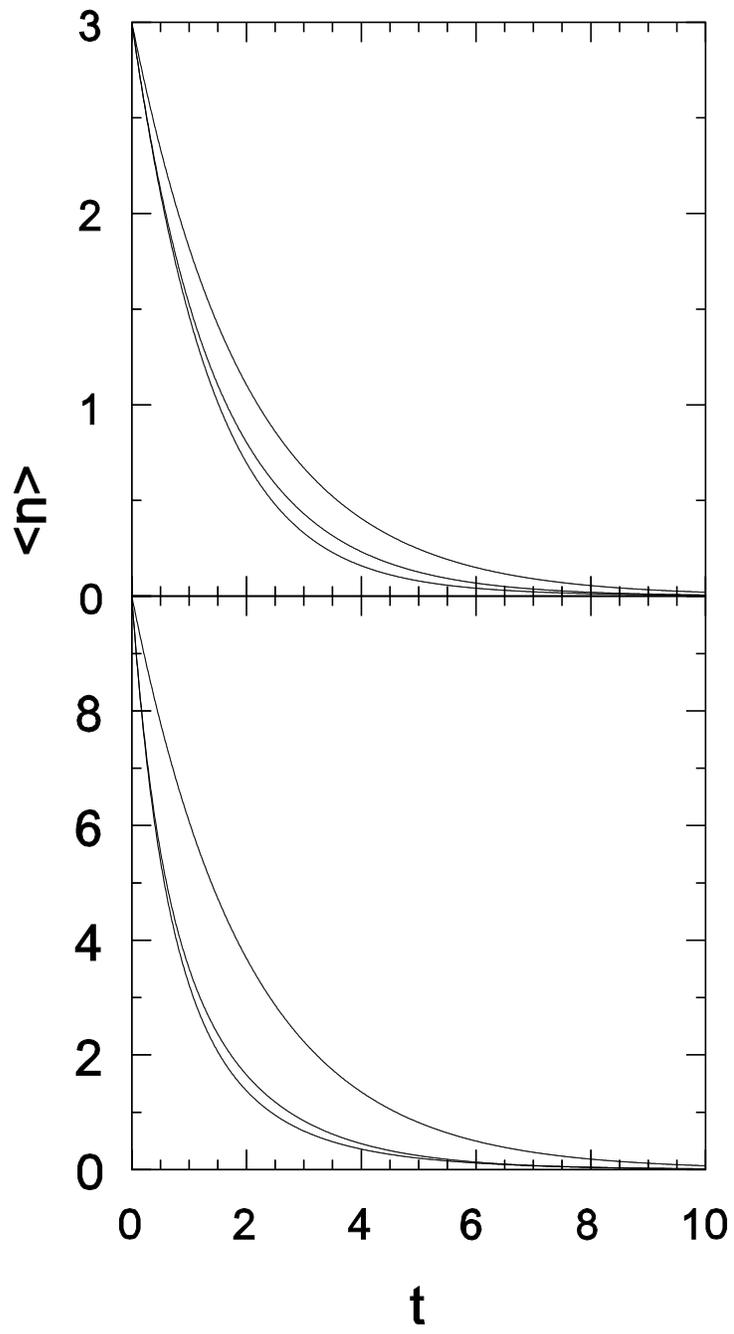
\end{document}